Transport Measurement of Landau level Gaps in Bilayer Graphene
J. Velasco Jr., Y. Lee, Z. Zhao, Lei Jing, P. Kratz, Marc Bockrath, and C. N. Lau[1]

**Abstract**
Landau level gaps are important parameters for understanding electronic interactions and symmetry-broken processes in bilayer graphene (BLG). Here we present transport spectroscopy measurements of LL gaps in double-gated suspended BLG with high mobilities in the quantum Hall regime. By using bias as a spectroscopic tool, we measure the gap $\Delta$ for the quantum Hall (QH) state at filling factor $\nu=\pm4$ and -2. The single-particle $\Delta_{\nu=4}$ scales linearly with magnetic field $B$ and is independent of the out-of-plane electric field $E_\perp$. For the symmetry-broken $\nu=-2$ state, the measured value of $\Delta_{\nu=-2}$ are ~1.1 meV/T and 0.17 meV/T for singly-gated geometry and dual-gated geometry at $E_\perp=0$, respectively. The difference between the two values arises from the $E_\perp$–dependence of $\Delta_{\nu=-2}$, suggesting that the $\nu=-2$ state is layer polarized. Our studies provide the first measurements of the gaps of the broken symmetry QH states in BLG with well-controlled $E_\perp$, and establish a robust method that can be implemented for studying similar states in other layered materials. .

The quantum Hall (QH) effect is a prototypical two-dimensional (2D) phenomenon that provides a rich platform for the study of many body physics, electronic interactions and symmetry breaking processes[1]. As one of the latest additions to the family of 2D materials, BLG hosts a plentitude of unusual QH phenomena, such as the Berry phase of $2\pi$[2], a state at filling factor $\nu=0$ with diverging Hall *and* longitudinal resistance[3, 45-10], quantum Hall ferromagnetism[11, 12], and electric-field driven transitions between quantum Hall states[5, 6]. In particular, the lowest Landau level (LL) is 8-fold degenerate, owing to the spin, sub-lattice and orbital degeneracies[12-16]. Ordering induced by electronic interactions that emerge from these competing symmetries, with or without external fields, has been a topic under intense investigation.

Recently, the advent of high quality BLG samples enabled observation of complete lifting of the 8-fold degeneracy in the lowest LL [3-6, 10, 17, 18], which arises from the spin, valley and orbital degrees of freedom. However, the exact nature of the symmetry-broken QH states ($\nu=0$, 1, 2 and 3) remains under intense debate[19-47], and very little is known about these states or the sizes of the LL gaps $\Delta$. In previous studies, $\Delta$ was measured by thermal activation of longitudinal resistivity of substrate-supported BLG devices[4], or by using scanned single-electron transistor on suspended samples[17]. However, gaps of the symmetry-broken states obtained from these studies differ by more than an order of magnitude.

Another salient fact is that the $\nu=\pm1$ and $\nu=\pm2$ QH states have broken spin and/or valley symmetries; since the valley and layer degrees of freedom are equivalent in the lowest Landau level, these QH states are expected to be sensitive to out-of-plane electric field $E_\perp$ that favors one valley/layer over another, hence their gaps should be $E_\perp$-dependent. Indeed, previous works[5, 6] have shown that these states are better resolved in the presence of $E_\perp$, suggesting a layer polarization component. However, to date all studies of LL gaps were performed on single-gated samples where $E_\perp$ is not independently controlled.

---
[1] Email address: lau@physics.ucr.edu

Here we present transport measurements of Landau level gaps in dual-gated suspended BLG by using source-drain bias $V$ as a spectroscopic tool in the quantum Hall regime. At a constant magnetic field $B$, plotting the device's two-terminal differential conductance $G=dI/dV$ as a function of source-drain bias $V$ and charge density $n$ yields a series of distinct diamond-shaped regions, which correspond to and evolve with quantum Hall plateaus. These diamonds arise from charge transport across graphene when the electrochemical potential in the edge states is aligned to or detuned from Landau levels in the bulk of the device, and yield information on the bulk gap and edge channel transport. The accuracy of the technique is established via measurements of the LL gaps for the $\nu=\pm 4$ state, which is independent of $E_\perp$ and scales linearly with $B$, in agreement with prior theoretical[12-16] and experimental[17, 48, 49] results. Using this technique, we measure the evolution of the gaps of the symmetry-broken $\nu=-2$ state as a function of $B$. At $E_\perp=0$, its gap $\Delta_{\nu=-2}$ is ~0.17 meV/T. In contrast, when only back gate is employed, $\Delta_{\nu=-2}$ is much larger, ~1.1 meV/T. This larger value, while in agreement with prior work on singly-gated devices[17], arises from sensitivity of $\Delta_{\nu=-2}$ to $E_\perp$, which is in turn induced by the single back gate and scales with $B$. Thus dual-gated devices are critical for accurate determination of gaps of symmetry-broken QH states in BLG. More generally, we demonstrate a simple and direct transport measurement of Landau level gaps that, unlike scanned probe measurements, is applicable to dual-gated samples, and can also be extended to study the gapped states of other atomic layers.

Double-gated suspended BLG devices (Fig. 1a) are fabricated using a multi-level lithographic technique[50, 51] and released from substrates by HF etching. The devices' field effect mobility values at $T=260$mK range from 40,000 to 120,000 cm$^2$/Vs. The presence of two gates allows independent adjustment of $n$ and $E_\perp$. Here we focus on a BLG device with mobility ~ 80,000 cm$^2$/Vs (similar data were observed in more than 5 devices). Fig. 1b displays the Landau fan data from a BLG device, plotting $G$ (color) at $V=0$ as a function of back gate voltage $V_{bg}$ (with the top gate disconnected or grounded) and $B$ for $0<B<8$T. The bands that radiate from $V_{bg}=0$ and $B=0$ correspond to quantum Hall plateaus. From the slopes of the plateaus, the back gate coupling ratio is estimated to be ~ 3.1 x 10$^{10}$ cm$^{-2}$V$^{-1}$. Line traces $G(V_{bg})$ at $B=1$T, 2T, 4T, 6T and 8T are shown in Fig. 1c. The 4-fold degenerate QH plateaus at $\nu=\pm 4$ and $\pm 8$ are resolved at $B=1$T; at higher $B$, symmetry-broken states at $\nu=-1$ and -2 are observed.

The above data are taken at source-drain bias $V=0$. Using $V$ as an additional variable, we plot $G(V, V_{bg})$ at constant $B$ (Fig. 2a). Interestingly, such a plot yields a striking series of diamonds, whose centers correspond exactly to the center of a QH conductance plateau. At larger $B$, as the Landau level gaps increase, the QH plateaus widen, and the sizes of the diamonds increase accordingly, thus suggesting that these diamonds are related to the QH states in BLG.

A close examination reveals that these diamonds consist of a series of zero-bias conductance valleys and peaks, whose positions are dependent on QH plateaus. For instance, at $B=1.3$T, the prominent diamond centered at $V_{bg}=-4$V (or $n\sim-1.3\times10^{11}$ cm$^{-2}$) corresponds to the center of $\nu=-4$ plateau, and a line trace $G(V)$ at this density displays a U-shaped conductance valley around zero bias (Fig. 2c). On the other hand, at $\nu\sim-3$ where *no* plateau is resolved, a zero bias conductance *peak* appear, corresponding to the *closing* of a diamond (Fig. 2d).

Such a series of diamonds strongly resemble the Coulomb blockade data from a quantum dot, *i.e.* transport across a device with quantized energy levels. Combined with the perfect correspondence between the diamonds and QH plateaus, we thus attribute the quantized levels to Landau levels, which are energetically separated by energy gaps $\Delta_\nu$. Our experimental observations can then be understood by considering transport of the device in the QH regime, where the measured conductance consists of contributions from both the bulk and the edge channels. When the zero-bias conductance is at the center of the QH plateau, BLG's Fermi level is pinned between the highest filled and the next unfilled LLs (Fig. 3a). Charges are carried by edge states, which are separated from the gapped bulk by a gap on the order of LL gap. Hence electrons are injected into the edge states and can tunnel into the bulk, yielding features that resemble tunneling spectroscopy[54]. Increasing bias raises the source's Fermi level, which eventually aligns with the next unfilled LL (Fig. 3b) and allows additional charge transport through the bulk, thereby leading to increased conductance. As a result, the device displays a conductance valley at $V=0$. Conversely, when $G$ is between the plateaus, the electrodes' Fermi level is aligned with the highest filled LL (Fig. 3c), thus allowing transport through the extended states in the bulk. Increasing bias detunes from the LL and disallows bulk transport at the Fermi level of the source contact, yielding lower conductance at large $V$, and thus an overall conductance peak at $V=0$. This model assumes non-equilibration of charges between the edge states and the bulk, which is reasonable considering the small dimension of the samples ($\sim$1-1.4 µm). We also note that a crucial component of the model is the ballistic transport of charges in the devices. Indeed, for devices with lower mobility, the diamond features are smeared or absent altogether.

Based on this simple model, we can spectroscopically resolve LL gaps by examining the $G(V)$ curves at the center of a QH plateau. The conductance at $V=0$ yields the edge state contribution, whereas the full width of the conductance valley yields $2\Delta$, where $\Delta$ is the gap between the filled and unfilled LLs. As a demonstration, we examine the $\nu=\pm 4$ states, whose gaps have been established experimentally[48, 49] and theoretically[12-14, 52, 53] to be $\Delta_{\nu=4} = \sqrt{2}\frac{\hbar eB}{m^*}$, where $h$ is Planck's constant, $e$ is electron charge, $m^*\sim 0.02 - 0.04 m_e$ is the effective mass of the charge carriers, and $m_e$ is electron's rest mass. To this end, we take $G(V)$ traces at $\nu=\pm 4$ and measure half-widths of the conductance valleys. The data points are taken at several different values of $B$, and for each $B$ value, at $E_\perp=0$ and $-14.4$ mV/nm. $E_\perp$ is calculated from $(n_{bg}-n_{tg})/2\varepsilon_0$, where $n_{bg}$ and $n_{tg}$ are charge density induced by back gate and top gate, respectively, and $\varepsilon_0$ is the permittivity of vacuum. The resultant data are shown in Fig. 3d. All the data points fall on a single straight line, independent of $E_\perp$. The slope of the best-fit line is 5.5 mV/T, thus yielding $m^*\approx 0.03 m_e$. Both the linear dependence on $B$ and the effective mass are in excellent agreement with prior experiments[17, 48, 49]. The lack of dependence of $\Delta_{\nu=\pm 4}$ on $E_\perp$ is also expected, since this is a single particle gap that does not involve layer polarization. Taken together, these results establish bias spectroscopy as a viable tool for measuring LL gaps.

We now focus on the state at filling factor $\nu=-2$, which arises from electronic interactions and has broken symmetries. We first perform bias spectroscopy with the top gate disconnected or grounded (both cases yield identical data). Fig. 4a plots $G(V,V_{bg})$ at $B$=3.5T. The $G(V)$ curve at $\nu=-2$ is shown as the red curve in Fig. 4c, displaying a well-resolved conductance valley at zero bias. The measured values of $\Delta_{\nu=-2}$ are plotted as a function of $B$ (red triangles) in Fig. 4d. The data points can be fitted to a straight line, with a slope ~1.1 meV/T, in very good agreement with prior work on singly-gated devices[17].

However, a significant drawback of singly-gated measurements is the inevitable presence of an interlayer electric field $E_\perp = ne/2\varepsilon_0$: for a given filling factor $\nu$ in magnetic field $B$, a single-gated BLG sheet experiences

$$E_\perp = \left(\frac{1}{2\varepsilon_0}\frac{e^2}{h}\right) B\nu \approx 2.2\ B\nu = 4.4\ \nu\ \text{(mV/nm)};$$

hence in these devices, $\Delta_{\nu=-2}$ is in fact measured at varying $E_\perp$ values, and does not truly characterize the state.

Taking advantage of the dual-gated geometry of our devices, we measure $\Delta_{\nu=-2}$ at $E_\perp$=0 by measuring $G(V,n)$ at different $B$. Fig. 4b displays such a plot at $B$=3.5T. Comparing with $G(V,V_{bg})$ data, the diamond at $\nu=-2$ is much diminished; the $G(V)$ traces display small conductance dips superimposed on a peak at zero bias (blue lines), suggesting only partial resolution of the symmetry-broken states. The measured $\Delta_{\nu=-2}$ values are shown as blue squares in Fig. 4d. They fall on a straight line, with a best-fit slope ~0.17 meV/T, which is almost an order of magnitude smaller than that measured on singly-gated devices. It is thus clear that the difference between the two data sets arises from the broken layer/valley symmetry of the $\nu=-2$ QH state, whose LL gap is likely to increase with applied $E_\perp$. Detailed study of evolution of this QH state with $E_\perp$ and $B$ will be explored in a future work[55].

Finally, we apply the bias spectroscopy to measure LL gap $\Delta_{\nu=-1}$. However, we are unable to resolve any discernible gap at $E_\perp$=0, thus putting an upper limit of 0.05 meV/T for this state. This value is again much smaller than those measured on singly-gated devices, suggesting that this $\nu=-1$ state has also a layer polarization component and an $E_\perp$–dependent gap. The measured values of $\Delta_{\nu=-2}$ and $\Delta_{\nu=-1}$ from this and prior works are summarized in Table 1.

In conclusion, we demonstrate that source-drain bias can be employed as a spectroscopic tool to measure the gaps of Landau levels in high mobility BLG. Using this technique, we find that the gaps of the $\nu=\pm 4$ states scale linearly with $B$, with an effective mass of 0.03 $m$, and is independent of electric field. In contrast, the gaps of the $\nu=-2$ state at finite $E_\perp$ exceeds those at $E_\perp$=0 by an order of magnitude, suggesting strong layer polarization in these states. Finally, dual-gated device geometry is crucial for accurate measurements of symmetry-broken LL gaps in BLG, since the electric field produced by a single gate is sufficient to induce layer polarization and preclude proper characterization this state at $E_\perp=0$.

We thank Fan Zhang and Yafis Barlas for helpful discussions. This work was supported in part by NSF CAREER DMR/0748910, NSF/1106358 and the FAME Center.

Fig. 1. (a). False-color SEM image of a dual-gated suspended BLG device. (b). $G(V_{bg},B)$ at zero bias in units of $e^2/h$. (c). Line trace $G(V_{bg})$ from (b) at $B$=1 (red), 2 (orange), 4 (green), 6 (blue) and 8T (purple), respectively.

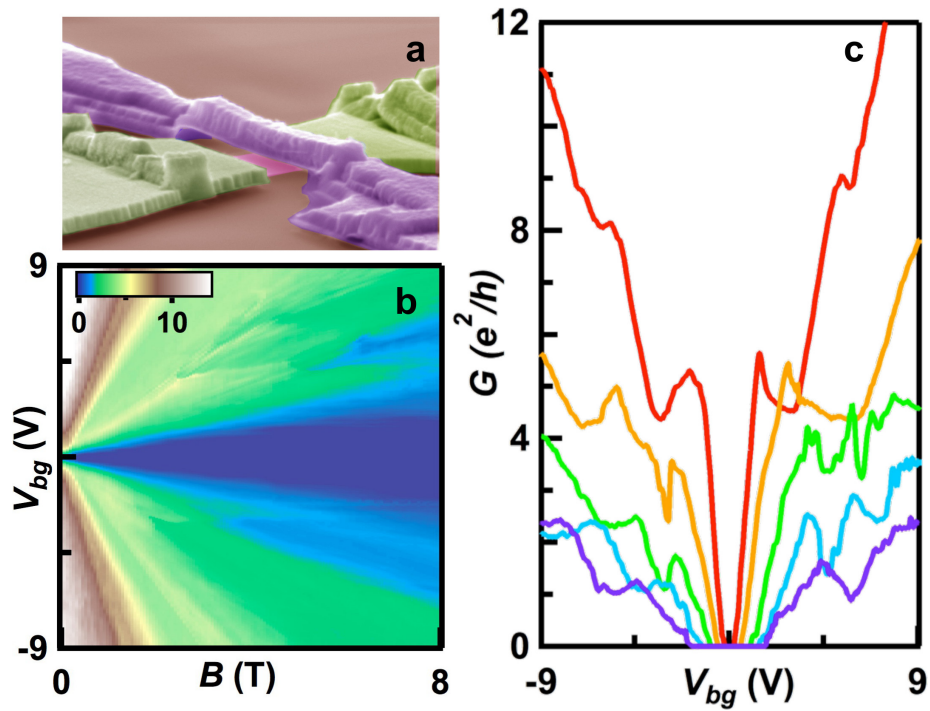

Fig. 2. (a). $G(V, V_{bg})$ in units of $e^2/h$ at $B=1.3$T. (b). Line trace $G(V_{bg})$ from (a) at $V=0$. The dotted lines indicate the correspondence between the diamond features in (c) and plateaus in (d). The numbers denote the filling factors. (c-d). Line traces $G(V)$ taken from (a) at $\nu=-4$ and $\nu=-3$, respectively.

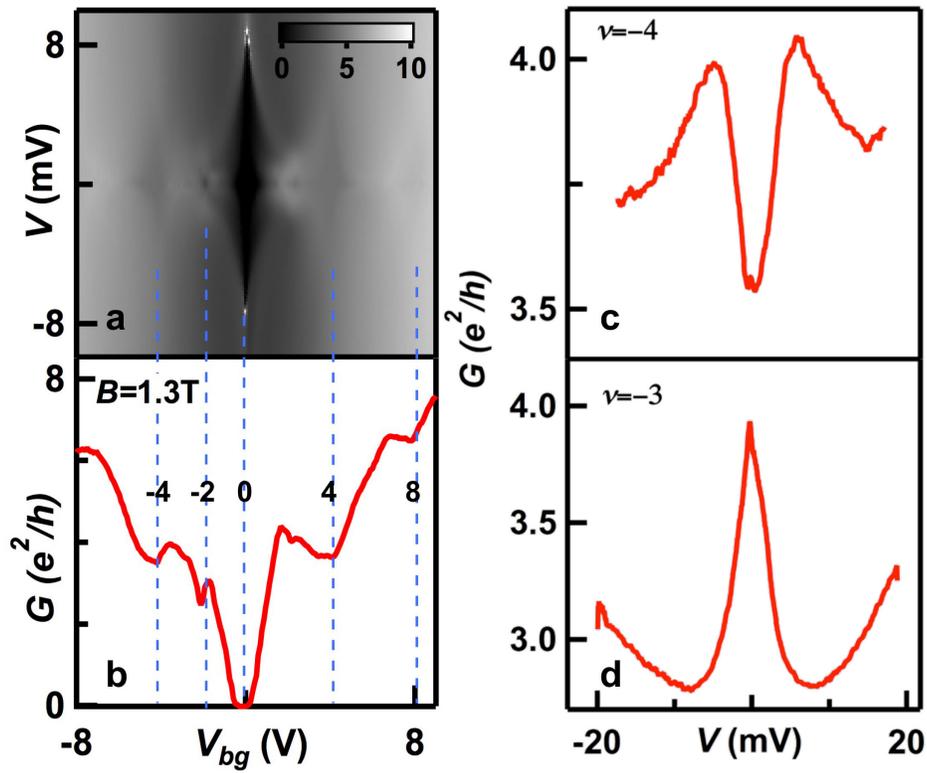

Fig. 3. Model of transport across the device in the quantum Hall regime and gap of the $v=4$ state. (a). At $V=0$, conductance is on a QH plateau. Fermi levels of the source (S) and drain (D) electrodes are located between the highest occupied Landau level and an unoccupied level. The bulk is gapped and transport occurs via edge states. (b). Similar to (a), except that a bias $V$ is applied between S and D. (c). At $V=0$, conductance is at the transition between QH plateaus. Here Fermi levels of the electrodes are aligned with the occupied Landau level. Transport occurs via the bulk. (d). Measured values of $\Delta_{v=\pm 4}$ as a function of $B$. The blue and red symbols indicate data taken at $v=4$ and $v=-4$, respectively; solid (hollow) symbols indicate data taken at $E_\perp=0$ and $E_\perp=-14.4$ mV/nm. The line is the best-fit to all the data points.

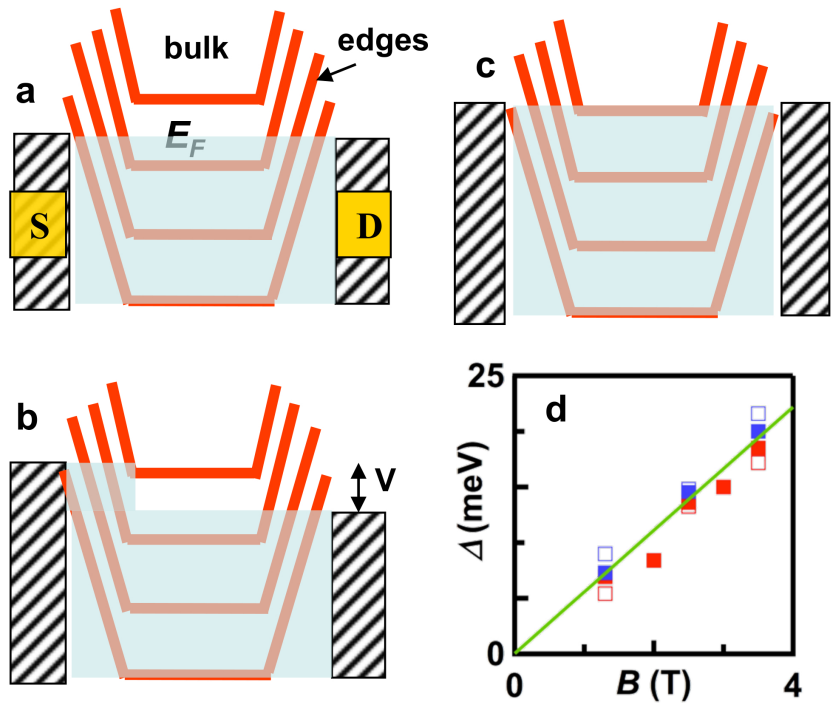

Fig. 4. (a). $G(V, \nu)$ data taken at $B$=3.5T with top gate disconnected or grounded. (b). $G(V, \nu)$ data taken at $B$=3.5T and $E_\perp$=0. (c). $G(V)$ line traces at $\nu$=-2 from (a) and (b). (d). Measured $\Delta_{\nu=-2}$ values. The red data points are taken with top gate disconnected or grounded, and blue data taken at $E_\perp$=0.

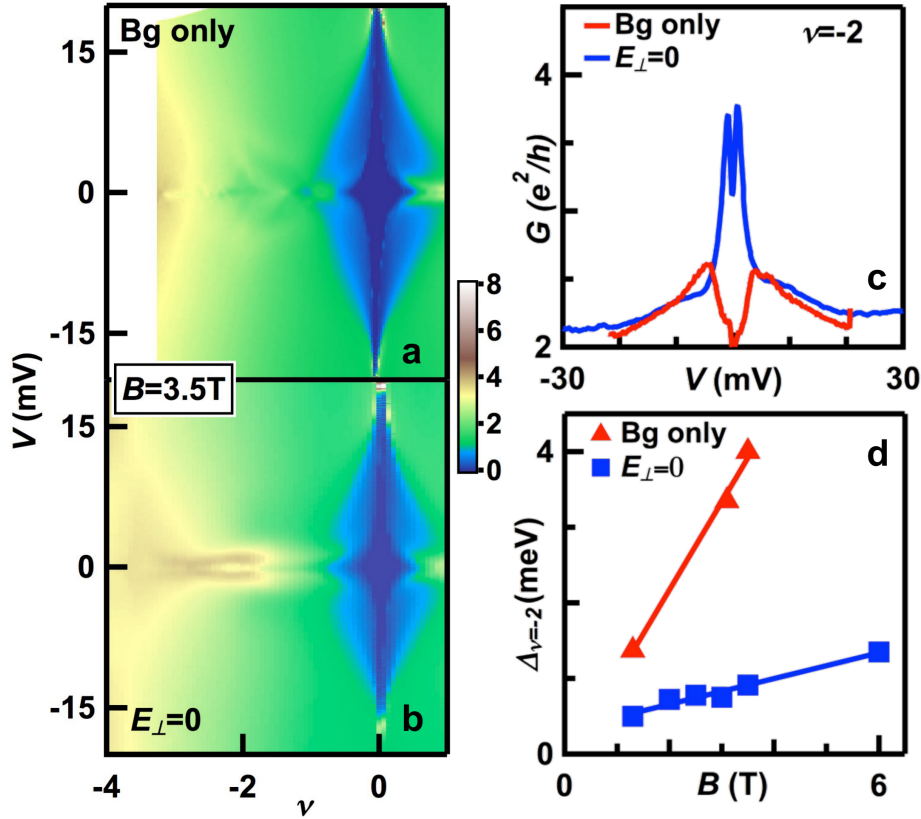

Table 1. Measured values of Landau level gaps for $\nu$=-2 and $\nu$=-1 states

|  |  | $\Delta_{\nu=-2}$ | $\Delta_{\nu=-1}$ |
|---|---|---|---|
| Ref.[4], single-gated on Si/SiO$_2$ |  | 0.05 meV/T | 0.01 meV/T |
| Ref.[17], single-gated, suspended |  | 1 meV/T | 0.1 meV/T |
| This work, suspended | Single-gated | 1.1 meV/T |  |
|  | $E_\perp$=0 | 0.17 meV/T | unresolved, <0.05 meV/T |